\documentclass[pra,groupedaddress,superscriptaddress,nofootinbib,floatfix,preprintnumbers,longbibliography,notitlepage,tightenlines,11pt]{revtex4-1}
\usepackage[paperwidth=210mm,paperheight=297mm,centering,hmargin=2.3cm,vmargin=2.3cm]{geometry}

\usepackage{amssymb,amsmath,graphicx,color,dsfont, mathrsfs 
,tikz-cd
}
\usepackage{bm}
\usepackage[export]{adjustbox}
\usepackage{braket}
\usepackage[colorlinks=true,citecolor=blue,linkcolor=blue
]{hyperref}
\usepackage{amsfonts}
\usepackage{amsthm}
\usepackage{braket}
\usepackage{enumerate}
\usepackage{adjustbox}
\usepackage{geometry}
\usepackage{array}
\usepackage{xfrac}
\usepackage{braket,subcaption}
\usepackage{url}
\usepackage{xcolor}
\usepackage{multirow}
\usepackage{rotating,booktabs}
\usepackage[verbose]{placeins}
\usepackage{ragged2e}
\usepackage[titletoc]{appendix}
\usepackage{soul}
\usepackage{boldline}

\newcommand{\kett}[1]{\left. \left| #1 \right\rangle \right\rangle}
\newcommand{\bbra}[1]{\left\langle \left\langle #1 \right| \right.}
\newcommand{\bbrakett}[1]{\left\langle \braket{#1} \right\rangle}

\begin{document}

\dimen\footins=5\baselineskip\relax

\preprint{IQuS@UW-21-116}

\title{Loop-string-hadron approach to SU(3) lattice Yang-Mills theory,   II: \\Operator representation for the trivalent vertex}

\author{Saurabh V. Kadam}
\email{ksaurabh@uw.edu}
\affiliation{
InQubator for Quantum Simulation (IQuS), Department of Physics, University of Washington, Seattle, WA 98195, USA}

\author{Aahiri Naskar}
 \email{p20230046@goa.bits-pilani.ac.in}
\affiliation{Department of Physics, BITS-Pilani,
K K Birla Goa Campus, Zuarinagar, Goa 403726, India.}%

\author{Indrakshi Raychowdhury}
 \email{indrakshir@goa.bits-pilani.ac.in}
\affiliation{Department of Physics, BITS-Pilani,
K K Birla Goa Campus, Zuarinagar, Goa 403726, India.}%
\affiliation{Center for Research in Quantum Information and Technology, Birla Institute of Technology and Science Pilani, K K Birla Goa Campus,  403726, India.}

\author{Jesse R.~Stryker}
\email{jstryker@lbl.gov}
\affiliation{
Physics Division, Lawrence Berkeley National Laboratory, Berkeley, CA 94720, USA.}

\date{\today}

\begin{abstract}
This work is the second installment of a series on the loop-string-hadron (LSH) approach to SU(3) lattice Yang-Mills theory.
Here, we present the infinite-dimensional matrix representation for arbitrary gauge-invariant operators at a trivalent vertex, which results in a standalone framework for computations that supersedes the underlying Schwinger-boson framework.
To that end, we present a partial summary of the commutation relations and use it to evaluate the result of applying any gauge-invariant operator on the LSH basis states introduced in Part I (\href{https://arxiv.org/abs/2407.19181}{arXiv:2407.19181}). Classical calculations in the LSH basis run significantly faster than equivalent calculations performed using Schwinger bosons. A companion code script is provided, which implements the derived formulas and aims to facilitate rapid progress towards Hamiltonian-based calculations of quantum chromodynamics. 
\end{abstract}

\maketitle 

\tableofcontents

\section{Introduction}
\noindent
\label{sec:int}

Dynamics of lattice gauge theories (LGTs) are of fundamental interest to explore and understand non-equilibrium phenomena and thermalization as well as for prediction of real-time evolution. This task, while being limited within the scope of conventional lattice quantum chromodynamics (QCD) calculations, is of increasing interest in the context of quantum computation and quantum simulation. Over more than the last decade, there has been significant progress in developing novel Hamiltonian frameworks~\cite{Carena:2022kpg, Banuls:2017ena, Raychowdhury:2019iki, Kadam:2022ipf, Pardo:2022hrp, Chandrasekharan:1996ih, Bauer:2021gek, Zohar:2014qma,Zohar:2018cwb,Zohar:2019ygc,Alexandru:2023qzd,Alexandru:2019nsa,Ji:2022qvr,Ji:2020kjk,Kavaki:2024ijd,DAndrea:2023qnr,Romiti:2023hbd,Zache:2023dko,Wiese:2021djl, Kaplan:2018vnj, Mathur:2015wba, Mathur:2023lky, Fontana:2024rux, Grabowska:2024emw,Hartung:2022hoz, Burbano:2024uvn, Muller:2023nnk,Ciavarella:2024fzw, Haase:2020kaj, Kane:2022ejm, Surace:2023qwo, Zache:2023cfj, Gonzalez-Cuadra:2022hxt,Illa:2024kmf, Zohar:2016iic, Lamm:2024jnl, Illa:2025dou, Illa:2025njz} and demonstrating quantum computing capabilities for the same
with the aim of computing dynamics of such complicated theories, starting with the simpler cases of the Schwinger model, quantum link models, and one-dimensional fully-gauge-fixed fermionic Hamiltonians for Abelian and non-Abelian gauge groups.
For recent reviews, see Refs.~\cite{Bauer:2022hpo,Bauer:2023qgm, DiMeglio:2023nsa, Halimeh:2025vvp}.

The starting point for any such calculation is a Hamiltonian framework, first proposed by Kogut and Susskind~\cite{Kogut:1974ag}, which comes with a description of basis vectors spanning the Hilbert space. However, such a Hilbert space suffers from having unphysical states due to the redundancies arising from the gauge symmetry. Thus, it becomes necessary to separate out the physical (i.e., gauge invariant) Hilbert space from the entire Hilbert space, which is often done by imposing the Gauss's law constraints on states. The Hamiltonian operator commutes with the gauge-symmetry constraints, and hence the dynamics remain confined only within the physical subspace. In the loop-string-hadron (LSH) formulation \cite{Raychowdhury:2019iki,Kadam:2022ipf,Kadam:2024ifg}, the non-commuting Gauss's law constraints are traded for Abelian constraints that could potentially reduce the cost of simulating non-Abelian LGTs, for both classical \cite{Davoudi:2020yln} and quantum \cite{Raychowdhury:2018osk,Davoudi:2022xmb} simulations.
The roadmap for simulating lattice Yang-Mills theory using the LSH formulation involves three foundational steps: first, characterizing the basis vectors that span the local physical subspace at each lattice site;
second, characterizing the local gauge-invariant operators and deriving their representation in the LSH basis;
and lastly, generalizing to the lattice by characterizing physical lattice states and expressing extended operators (like Wilson loops and the Hamiltonian) in terms of local LSH operators.
The present work addresses the second step for pure SU(3) LGT at a trivalent vertex, which is the building block of a LGT in more than one spatial dimension.

For the case of pure gauge theories, the gauge-invariant Hilbert space is created by manifestly gauge-invariant Wilson loop operators. Computing the dynamics in terms of Wilson loops is nontrivial as the Wilson loops define an overcomplete and non-local basis. Schwinger bosons provide a partial solution for this by solving the problem of non-locality in terms of site-local gauge-invariant operators and states, which are referred to as `local loops'. For the case of SU(2) LGT, various approaches can alleviate the issue of an overcomplete basis \cite{watson1994solution, loll1993yang, mathur2006loop, anishetty20142}. The loop-string-hadron (LSH) framework for SU(2) LGTs, derived from Schwinger bosons, provides an alternative for obtaining a local and orthonormal gauge-invariant loop basis in any number of spatial dimensions~\cite{Raychowdhury:2019iki}. This allows computing dynamics for SU(2) in terms of strictly gauge-invariant degrees of freedom and is being explored in the context of quantum simulating SU(2) LGTs \cite{Raychowdhury:2018osk, Dasgupta:2020itb, Davoudi:2020yln, Davoudi:2022xmb, Mathew:2025fim}. 

However, the generic solution of constructing a complete and orthonormal basis for SU(3) gauge theory continues to pose a technical hurdle~\cite{Littlewood1934GroupCA,knuth1970permutations,GopalkrishnaGadiyar:1991ah,2023CMaPh.400..179C,Balaji:2025yua}.
The lessons from the SU(2) LSH framework guide us to tackle the problem at the level of its individual building blocks. While the point-splitting scheme for SU(2) resolves all issues at four- and six-point vertices \cite{Raychowdhury:2019iki}, the same procedure does not by itself resolve all issues at a vertex for SU(3) \cite{Kadam:2024ifg}. It continues to be true that a trivalent vertex can be a building block of multidimensional lattices, but previous studies have shown that there is more to understand for building a SU(3) gauge theory Hilbert space for even a single trivalent vertex.
One way of dealing with the trivalent vertex would be to describe the construction of an orthogonalized Hilbert space algorithmically, see e.g. Ref.~\cite{Balaji:2025yua}.
But ideally one would have closed-form, analytic expressions for the states so that matrix elements of operators can be worked out once and for all, without the need for orthogonalization on a case-by-case basis.

In Part I \cite{Kadam:2024ifg} of the current series of systematic developments, we reported on the construction of the gauge-invariant Hilbert space of a trivalent vertex, its construction from specialized combinations of Schwinger-boson multiplets (also known as prepotentials), issues of orthogonality and counting degrees of freedom, and finally, a proposal for orthogonalization that defines the basis implicitly.
The current article, Part II in the series, provides the matrix elements of the gauge-invariant operators that are the building blocks for constructing the SU(3) LGT Hamiltonian in the LSH framework.
Importantly, in Part I, most results were calculated in terms of their Schwinger-boson definitions. While irreducible Schwinger bosons (ISBs) remain the underlying foundation for the SU(3) LSH formulation, the focus of this work is to
present a framework for doing calculations at a vertex exclusively in terms of LSH quantum numbers, eliminating the need to constantly decompose operations into their Schwinger-boson definitions.
We further note that the basis used here is the explicit nonorthogonal basis presented in Part I, however, we expect the results presented here are a prerequisite for the LSH formulation in terms of orthogonal degrees of freedom.
In the upcoming papers, we will complete the story by focusing on applying the results of this work beyond a single vertex -- to the entire lattice, where the additional generalizations needed in going from the SU(2) LSH formulation to the SU(3) LSH formulation are less substantial, but are yet to be explored.

The plan of this article is as follows. The stage is set with defining the notation and conventions in Sec.~\ref{sec: notations}. The main results are then summarized in Sec.~\ref{sec: resultspart2}. 
Section \ref{sec: commutators} presents a number of algebraic identities that are useful for deriving the matrix-representation formulas that follow in Sec.~\ref{sec: resultsanalytic}, while Sec.~\ref{sec: resultscode} serves to introduce the companion Mathematica notebook that applies our matrix-representation formulas and works exclusively in terms of LSH degrees of freedom. The derivations of the matrix representations reported in Sec.~\ref{sec: resultsanalytic} are explained in the later Sec.~\ref{sec: derivationspart2}, and finally a summarizing discussion and outlook are provided in Sec.~\ref{sec: discussionpart2}.

\section{\label{sec: notations}Notation and conventions}
\newcommand{\qnums}{\textbf{q}}
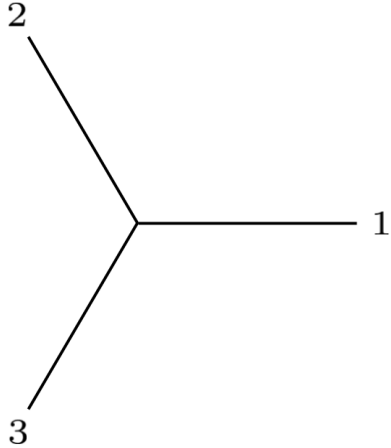
\emph{Labels and indices.}---We consider a trivalent vertex with legs labeled as $1$, $2$, and $3$. In this Part II manuscript, we always use lowercase Roman letters $i,j,k$ as variables to refer to vertex legs\footnote{Note that in Part I we used capital $I,J,K$ as variables that refer to vertex legs. Switching to $i,j,k$ reduces the width of some long formulas. Furthermore, we have repurposed $J$ and $K$ for operators in the present work.}.
In Sec.~\ref{sec: methods} and later, there is a potential for confusion between vertex-leg labels and the color index for SU(3) triplets or antitriplets because in both cases the labels take values 1, 2, and 3; for clarity, we use lowercase Greek letters exclusively for color indices.

Unless noted otherwise, when reading equations, one should assume $(i,j,k)$ is a permutation of the legs (1,2,3).
That is, $i$, $j$, and $k$ have fixed distinct values in each identity.
The Einstein summation convention is never applied for leg labels.

\emph{Gauge-singlet operators.}---
At the trivalent vertex, there is a great variety of gauge-singlet operators that are to be understood.
The operators one may form at the trivalent vertex may involve one, two, or all three legs.
The operators of principle interest in this work are the so-called ``corner'' operators \cite{Burbano:2024uvn}, involving two legs of the vertex $(i,j)$.
This is because the corner operators are used to construct Wilson loop operators on extended lattices \cite{Anishetty:2009nh,Raychowdhury:2019iki}.
Here the notation for the gauge-singlet operators shall be introduced, while their original constructions in terms of the ISB modes will be discussed in \ref{sec:isb}.

The simplest operators are those defined on a single leg, the irrep quantum numbers $P_i$ and $Q_i$ which are Hermitian operators with integer eigenvalues 0, 1, 2, $\ldots$.
We also make frequent use of the derived operator
\begin{align}
    \label{eq: Fi definition}
    F_i &\equiv (P_i+Q_i+2)^{-1}.
\end{align}
The creation operators that define the gauge-invariant Hilbert space are denoted by $L_{ij}^\dagger$, $T_A^\dagger$, and $T_B^\dagger$, numbering eight in total.
The $L_{ij}^\dagger$ operators, along with their annihilation counterparts $L_{ij}$, are examples of corner operators.
The other corner operators to be focused on are denoted
\begin{align}
    N_{ij}=N_{ji}^\dagger, \quad 
    M_{ij}=M_{ji}^\dagger, \quad 
    J_{ij}, \quad
    J_{ij}^\dagger, \quad
    K_{ij}, \quad
    \mathrm{and} \quad K_{ij}^\dagger.
\end{align}
The operators $T_A^\dagger$ and $T_B^\dagger$, along with their annihilation counterparts $T_A$ and $T_B$, are examples of operators that involve all three vertex legs.
It turns out that $T_A^\dagger$, $T_B^\dagger$, $T_A$, and $T_B$ are not independent; they can all be obtained as commutators of corner operators.
Other examples of trivalent operators may be cataloged as
\begin{align}
\label{eq: other trilinear operators}
\epsilon A^\dagger_i A^\dagger_j B_k, \quad 
\epsilon B^\dagger_i B^\dagger_j A_k, \quad
\epsilon B^\dagger_i A_j A_k = - (\epsilon A^\dagger_k A^\dagger_j B_i)^\dagger, \quad
\mathrm{and} \quad 
\epsilon A^\dagger_i B_j B_k = - (\epsilon B^\dagger_k B^\dagger_j A_i)^\dagger.
\end{align}
Unlike the other operators, the trivalent operators in \eqref{eq: other trilinear operators} have been denoted using an abbreviated form of their ISB constructions instead of being assigned abstract symbols of their own.
There are two reasons for this:
(i) Like $T_A$ and $T_B$, these operators can always be reexpressed in terms of corner operators. (See \eqref{eq: ++- corner operator identities} and \eqref{eq: +-- corner operator identities}.)
(ii) At the time of writing, we do not know of a physical use case for these operators. In particular, they are necessary neither to construct the Hilbert space nor to express (Wilson loop) operators in the Hamiltonian.

\emph{Basis states.}---Associated to the vertex is a basis of gauge-invariant states $\kett{\mathbf{q}}$, characterized by seven integer quantum numbers:
\begin{align}
    \label{eq: vertex basis states}
    \kett{\mathbf{q}} &= \kett{\ell_{12}\, , \ell_{23}\, , \ell_{31}\, , \ell_{21}\, , \ell_{32}\, ,\ell_{13}, t } ,
\end{align}
where 
\begin{align}
    \ell_{ij} &\in \{ 0, \ 1, \ 2, \ 3, \, \ldots \} \quad \mathrm{and} \quad t \in \{ 0, \ \pm 1, \ \pm 2 , \ \pm 3, \ \ldots \} .
\end{align}
These basis states, used throughout this work, are the so-called ``naive LSH basis'' introduced in Part I \cite{Kadam:2024ifg}.
The double-angle bras and kets, $\bbra{*}$ and $\kett{*}$, are used for the basis states to emphasize that they are generally not normalized to unit length, that is, the basis is unnormalized.
In addition, the basis states are not completely orthogonalized, hence we interchangeably refer to this basis as the ``nonorthogonal'' basis.
For further discussions on orthogonality and overlaps, we refer the reader to Part I.

As explained in Part I, the basis kets are ultimately defined in terms of ISBs, and this definition is reviewed in Sec.~\ref{sec: methods}; however the significance of this Part II article is that our results make it possible to use the states \eqref{eq: vertex basis states} without reverting to their underlying Schwinger-boson definitions. This includes the ability to calculate norms and overlaps.

\emph{Shorthand expressions.}---
In writing, it is convenient to introduce some shorthand expressions and abuses of notation:
\begin{itemize}
    \item A completely omitted quantum number in a basis ket is understood to be equal to zero. For example, $\kett{\ell_{12}, t}$ is a shorthand notation for the states $\kett{\ell_{12} , 0 , 0 , 0 , 0 , 0 ,t }$.
    \item The ordering of the quantum numbers in a basis ket is sometimes not adhered to. For example, $\kett{\ell_{ij},\ell_{ji}}$ refers to states $\kett{\ell_{12} , 0 , 0, \ell_{21} , 0 , 0, t }$ for both $(i,j)=(1,2)$ and $(i,j)=(2,1)$.
    \item In the Results (Sec.~\ref{sec: resultspart2}), on the right-hand sides of equations, we may suppress some quantum numbers in kets by using `$\cdots$' if they are unchanged relative to the ket $\kett{\qnums}$, for example, see Eq.~\eqref{eq: Lijdagger action}.
    \item For compactness, it can be convenient to use the symmetrized variables
\begin{align}
    \sigma_{ij} &\equiv \ell_{ij} + \ell_{ji}.
\end{align}
\end{itemize}
Of course, in the companion Mathematica code, none of the shorthand conventions for kets have any role and the quantum numbers always match the format of \eqref{eq: vertex basis states}.

\section{\label{sec: resultspart2}Results}
The gauge-invariant operators at the vertex must be understood in order to 
construct a lattice Hamiltonian matrix,
express the observables,
and normalize the states.
The operators are formally defined in terms of Schwinger bosons; however, as remarked previously, it is tedious and computationally expensive to work with them.
To lift the calculations from their Schwinger-boson definitions to a more compact, LSH-basis representation, it is sufficient to know the matrix representation of any necessary local operator in this basis.
What we mean by having a matrix representation for some linear operator $\mathcal{O}$ associated with the vertex is that, for a given basis ket $\kett{\mathbf{q}}$, the state
\begin{align}
    \mathcal{O} \kett{ \mathbf{q} } &= \sum_{ \mathbf{q'} } [ \mathcal{O} ]_{ \mathbf{q'}, \mathbf{q} } \kett{ \mathbf{q'} }
\end{align}
has coefficients $[ \mathcal{O} ]_{ \mathbf{q'}, \mathbf{q}}$ that are reported in, or trivially derivable from, the results below. 
Furnishing the coefficients $[ \mathcal{O} ]_{ \mathbf{q'}, \mathbf{q} }$ is the main object of this article, and our method of  calculation requires a number of commutator identities to accomplish that.
In Sec.~\ref{sec: commutators}, a number of helpful commutators are summarized, then the matrix representations of operators are reported in Sec.~\ref{sec: resultsanalytic}.
Section \ref{sec: resultscode} serves to introduce the companion Mathematica notebook that implements the formulas in Sec.~\ref{sec: resultsanalytic}.

\subsection{Algebra of operators}
\label{sec: commutators}
Below is an partial summary of the commutators of gauge-singlet operators at the trivalent vertex introduced in Sec.~\ref{sec: notations}.
The commutators themselves may be derived by using elementary properties of the irreducible Schwinger bosons given in Part I and restated in Sec.~\ref{sec:isb}.
In general, evaluating the commutators requires only methods that have been discussed or used in earlier works, and therefore we do not present their proofs.

First, all of the pure-creation singlet operators commute with each other:
\begin{subequations}
    \label{eq: pure-creation commutators}
    \begin{align}
        \left[ L_{ij}^\dagger  , L_{i' j'}^\dagger \right] &= 0 \text{ for any corners $(i,j)$ and $(i',j')$}, \\
        \left[ L_{ij}^\dagger  , T_A^\dagger \right] = \left[ L_{ij}^\dagger  , T_B^\dagger \right] &= 0 .
    \end{align}
\end{subequations}

The next simplest commutators are those expressing how the various operators raise or lower the $P_i$ and $Q_i$ quantum numbers, which have been collected into Table \ref{tab: Pi and Qi algebra}.
For example, the column under $[*,N_{ij}]$ indicates that $N_{12}$ raises $P_1$ by one unit, lowers $P_2$ by one unit, and leaves $P_3$, $Q_1$, $Q_2$, and $Q_3$ invariant.
Some omitted formulas, such as $[P_{i},T_A]$ can be obtained by Hermitian conjugation.
Note that the rules for commuting $F_i$ operators through the various gauge-singlet operators are inherited from raising and lowering properties; for example, $F_j N_{ij} T_A^\dagger = N_{ij} (P_j+Q_j+1)^{-1} T_A^\dagger = N_{ij} T_A^\dagger F_j$.
\begin{table*}[!tbhp]
  \renewcommand{\arraystretch}{1.2}
  \centering
    \setlength{\tabcolsep}{1pt}
    \resizebox{\textwidth}{!}{%
   \begin{tabular}{c V{3} cc V{3} ccc V{3} cccc}
     \multicolumn{1}{l}{$\quad$ }\\
&$[*, P_{i} ]$&$[*, Q_{i} ]$&$[*, L_{ij}^\dagger ]$&$[*, T_A^\dagger ]$&$[*, T_B^\dagger ]$&$[*, N_{ij} ]$&$[*, M_{ij} ]$&$[*, J_{ij}^\dagger ]$&$[*, K_{ij}^\dagger ]$ \\
\hlineB{3}
$[P_{i'} , * ]$&$ 0 $&$ 0 $&$ \delta_{i'i} L_{ij}^\dagger $&$T_A^\dagger$&$ 0 $&$ (\delta_{i'i}-\delta_{i'j})N_{ij} $&$ 0 $&$ (\delta_{i'i}+\delta_{i'j})J_{ij}^\dagger $&$ -\delta_{i'j}K_{ij}^\dagger $ \\
$[Q_{i'} , * ]$&$ 0 $&$ 0 $&$ \delta_{i'j}L_{ij}^\dagger $&$ 0 $&$ T_B^\dagger $&$ 0 $&$ (\delta_{i'i}-\delta_{i'j})M_{ij} $&$ -\delta_{i'j}J_{ij}^\dagger $&$ (\delta_{i'i}+\delta_{i'j})K_{ij}^\dagger $ \\
 \end{tabular}
 }
 \caption{\label{tab: Pi and Qi algebra} Commutators involving the number operators $P_i$ and $Q_i$. In these formulas, $i$, $j$, and $i'$ can each be any leg of the vertex as long as $i\neq j$.}
\end{table*}

Earlier, it was remarked that various trivalent operators are expressible in terms of corner operators.
Among others, one has the relations
\begin{alignat}{4}
    \label{eq: TA and TB operator identities}
    T_A &= \epsilon_{ijk} [ L_{kj}, J_{ij} ], & T_B &= \epsilon_{ijk} [ L_{jk}, K_{ij} ] , \\
    \label{eq: ++- corner operator identities}
    \epsilon A_i^\dagger A_j^\dagger B_k &= [ J^\dagger_{ij}, M_{jk}] = [M_{ik}, J_{ji}^\dagger], \quad & \epsilon B_i^\dagger B_j^\dagger A_k &= [ K^\dagger_{ij}, N_{jk}] = [N_{ik}, K_{ji}^\dagger] , \\
    \label{eq: +-- corner operator identities}
    \epsilon A_i^\dagger B_j B_k &= [ J_{ij}^\dagger, L_{jk} ] = [L_{kj}, J_{ik}^\dagger ], & \epsilon B_i^\dagger A_j A_k &= [ K_{ij}^\dagger, L_{kj} ] =[L_{jk},K_{ik}^\dagger].
\end{alignat}

Lastly, the bulk of the commutators to be used in Sec.~\ref{sec: methods} involve a corner operator and a pure-creation operator.
For the $J^\dagger_{ij}$ and $K^\dagger_{ij}$ corner operators, one has
\begin{subequations}
    \label{eq: JDagger and KDagger Commutators combined}
    \begin{alignat}{4}
        \left[ J^\dagger_{ij} , L^\dagger_{ij} \right]  &= 0, & 
        \left[ K^\dagger_{ij} , L^\dagger_{ij} \right]  &= 0, \\
        \left[ J^\dagger_{ij} , L^\dagger_{jk} \right]  &= 0, &
        \left[ K^\dagger_{ij} , L^\dagger_{jk} \right]  &= \epsilon_{ijk} T_B^\dagger, \\
        \left[ J^\dagger_{ij} , L^\dagger_{ki} \right]  &= 0, &
        \left[ K^\dagger_{ij} , L^\dagger_{ki} \right]  &= 0, \\
        \left[ J^\dagger_{ij} , L^\dagger_{ji} \right]  &= 0, &
        \left[ K^\dagger_{ij} , L^\dagger_{ji} \right]  &= 0, \\ 
        \left[ J^\dagger_{ij} , L^\dagger_{kj} \right]  &= \epsilon_{ijk} T_A^\dagger, &
        \left[ K^\dagger_{ij} , L^\dagger_{kj} \right]  &= 0, \\
        \left[ J^\dagger_{ij} , L^\dagger_{ik} \right]  &= 0, &
        \left[ K^\dagger_{ij} , L^\dagger_{ik} \right]  &= 0, \\
        \left[ J^\dagger_{ij} , T_A^\dagger    \right]  &= 0, &
        \left[ K^\dagger_{ij} , T_A^\dagger    \right]  &= \epsilon_{ijk} L^\dagger_{ki}L^\dagger_{ij}, \\
        \left[ J^\dagger_{ij} , T_B^\dagger    \right]  &= \epsilon_{ijk} L^\dagger_{ik}L^\dagger_{ji}, \qquad &
        \left[ K^\dagger_{ij} , T_B^\dagger    \right]  &= 0.
    \end{alignat}
\end{subequations}
For the $N_{ij}$ and $M_{ij}$ corner operators, they are
\begin{subequations}
    \label{eq: N and M Commutators combined}
    \begin{alignat}{4}
        \left[ N_{ij} ,  L^\dagger_{ij} \right] &= -F_jL^\dagger_{ij}N_{ij},  &
        \left[ M_{ij} ,  L^\dagger_{ij} \right] &= -F_jL^\dagger_{ji}N_{ij}, \\
        \left[ N_{ij} ,  L^\dagger_{jk} \right] &= L^\dagger_{ik}-F_jL^\dagger_{ij}M_{kj}, &
        \left[ M_{ij} ,  L^\dagger_{jk} \right] &= -F_jL^\dagger_{ji}M_{kj}, \\
        \left[ N_{ij} ,  L^\dagger_{ki} \right] &= 0, &
        \left[ M_{ij} ,  L^\dagger_{ki} \right] &= 0, \\
        \left[ N_{ij} ,  L^\dagger_{ji} \right] &= -F_jL^\dagger_{ij}M_{ij}, &
        \left[ M_{ij} ,  L^\dagger_{ji} \right] &= -F_jL^\dagger_{ji}M_{ij},  \\
        \left[ N_{ij} ,  L^\dagger_{kj} \right] &= -F_jL^\dagger_{ij}N_{kj}, &
        \left[ M_{ij} ,  L^\dagger_{kj} \right] &= L^\dagger_{ki}-F_jL^\dagger_{ji}N_{kj}, \\
        \left[ N_{ij} ,  L^\dagger_{ik} \right] &= 0, &
        \left[ M_{ij} ,  L^\dagger_{ik} \right] &= 0, \\
        \left[ N_{ij} ,  T_A^\dagger    \right] &= \epsilon_{ijk} F_j L_{ij}^\dagger ( \epsilon{A^\dagger_i}{A^\dagger_k}{B_j} ), \quad &
        \left[ M_{ij} ,  T_A^\dagger    \right] &= \epsilon_{ijk} F_j L^\dagger_{ji}\,(\epsilon{A^\dagger_i}{A^\dagger_k}{B_j}), \\
        \left[ N_{ij} ,  T_B^\dagger    \right] &= \epsilon_{ijk} F_j L^\dagger_{ij} ( \epsilon{B^\dagger_i}{B^\dagger_k}{A_j} ), &
        \left[ M_{ij} ,  T_B^\dagger    \right] &= \epsilon_{ijk} F_j L_{ji}^\dagger ( \epsilon{B^\dagger_i}{B^\dagger_k}{A_j} ).
    \end{alignat}
\end{subequations}
For the pure-annihilation operators $L_{ij}$, one has
\begin{subequations}
    \label{eq: L Commutators combined}
    \begin{align}
        \left[ L_{ij} ,  L^\dagger_{ij} \right] &= 3+P_i+Q_j-F_iM_{ij}M_{ji}-F_jN_{ji}N_{ij}-F_iF_jL^\dagger_{ji}L_{ji}, \\
        \left[ L_{ij} ,  L^\dagger_{jk} \right] &= -F_j N_{ji}M_{kj}, \\
        \left[ L_{ij} ,  L^\dagger_{ki} \right] &= -F_i M_{ij}N_{ki}, \\
        \left[ L_{ij} ,  L^\dagger_{ji} \right] &= -F_i M_{ij}N_{ji}-F_jN_{ji}M_{ij}-F_iF_jL^\dagger_{ji}L_{ij},\\
        \left[ L_{ij} ,  L^\dagger_{kj} \right] &= N_{ki}-F_j N_{ji}N_{kj}, \\
        \left[ L_{ij} ,  L^\dagger_{ik} \right] &= M_{kj}-F_i M_{ij}M_{ki}, \\
        \left[ L_{ij} ,  T_A^\dagger \right] &= \epsilon_{ijk} \bigl( - J^\dagger_{kj} - F_i M_{ij} ( \epsilon A_j^\dagger A_k^\dagger B_i ) - F_j N_{ji} (\epsilon A_k^\dagger A_i^\dagger B_j ) - F_i F_j L_{ji}^\dagger (\epsilon A_k^\dagger B_i B_j ) \bigr),\\
        \left[ L_{ij} ,  T_B^\dagger \right] &= \epsilon_{ijk} \bigl( K^\dagger_{ki} + F_j N_{ji} ( \epsilon B_i^\dagger B_k^\dagger A_j ) + F_i M_{ij} (\epsilon B_k^\dagger B_j^\dagger A_i ) + F_i F_j L_{ji}^\dagger (\epsilon B_k^\dagger A_j A_i ) \bigr).
    \end{align}
\end{subequations}
And for the $J_{ij}$ and $K_{ij}$ corner operators, they are
\begin{subequations}
    \label{eq: J and K commutators combined}
    \begin{alignat}{4}
        \left[ J_{ij} ,  L^\dagger_{ij} \right] &= F_i K^\dagger_{ij} M_{ji}, &
        \left[ K_{ij} ,  L^\dagger_{ij} \right] &= F_i J^\dagger_{ij} M_{ji} - J^\dagger_{ji}, \\
        \left[ J_{ij} ,  L^\dagger_{jk} \right] &= \epsilon B^\dagger_kB^\dagger_jA_i, &
        \left[ K_{ij} ,  L^\dagger_{jk} \right] &= 0, \\
        \left[ J_{ij} ,  L^\dagger_{ki} \right] &= F_i K^\dagger_{ij} N_{ki}, &
        \left[ K_{ij} ,  L^\dagger_{ki} \right] &= -J^\dagger_{kj} + F_i J^\dagger_{ij} N_{ki}, \\
        \left[ J_{ij} ,  L^\dagger_{ji} \right] &= F_i K^\dagger_{ij} N_{ji} - K^\dagger_{ji}, &
        \left[ K_{ij} ,  L^\dagger_{ji} \right] &= F_i J^\dagger_{ij} N_{ji}, \\
        \left[ J_{ij} ,  L^\dagger_{kj} \right] &= 0, &
        \left[ K_{ij} ,  L^\dagger_{kj} \right] &= \epsilon A^\dagger_k A^\dagger_j B_i , \\
        \left[ J_{ij} ,  L^\dagger_{ik} \right] &= -K^\dagger_{kj} + F_i K^\dagger_{ij} M_{ki}, &
        \left[ K_{ij} ,  L^\dagger_{ik} \right] &= F_i J^\dagger_{ij} M_{ki} , \\
        \left[ J_{ij} ,  T_A^\dagger    \right] &= \epsilon_{ijk} \bigl( L^\dagger_{kj} (P_i+P_j+2) - L^\dagger_{ij}N_{ki} & - F_i K^\dagger_{ij} (\epsilon A^\dagger_k& A^\dagger_jB_i) \bigr), \nonumber \\
        \left[ K_{ij} ,  T_A^\dagger    \right] &= -\epsilon_{ijk} F_i J^\dagger_{ij}(\epsilon A^\dagger_kA^\dagger_jB_i), & \\
        \left[ J_{ij} ,  T_B^\dagger    \right] &= -\epsilon_{ijk} F_i K^\dagger_{ij}(\epsilon B^\dagger_kB^\dagger_jA_i), \nonumber \\
        \left[ K_{ij} ,  T_B^\dagger    \right] &= \epsilon_{ijk} \bigl( L^\dagger_{jk} (Q_i+Q_j+2)-L^\dagger_{ji}M_{ki} & - F_i J^\dagger_{ij}(\epsilon B^\dagger_k & B^\dagger_jA_i) \bigr) .
    \end{alignat}
\end{subequations}

\subsection{\label{sec: resultsanalytic}Matrix representation of operators}

Below we consider a subset of gauge-singlet operators that is sufficient for expressing local dynamics.
Explicit formulas are given for applying any one of these operators on an arbitrary LSH basis ket, which is to say their matrix representations are given.

The simplest operators to express are the diagonal $P_i$ and $Q_i$ irrep quantum numbers, the functions $F_i = (2+P_i+Q_i)^{-1}$, and the pure creation operators $L^\dagger_{ij}$ and $T^\dagger_{A/B}$. They have the following matrix representations.
\begin{align}
P_i \kett{ \qnums } &= \bigl(\ell_{ij} + \ell_{ik} + |t| \, \theta(  t) \bigr) \kett{ \qnums } ,\\
Q_i \kett{ \qnums } &= \bigl(\ell_{ji} + \ell_{ki} + |t| \, \theta( -t) \bigr) \kett{ \qnums } ,\\
F_i \kett{ \qnums } &= (2+\sigma_{ij}+\sigma_{ik}+|t|)^{-1} \kett{ \qnums },\\
L_{ij}^\dagger \kett{ \qnums } &= \kett{ \ell_{ij}+1 , \cdots } , \label{eq: Lijdagger action} \\
T_A^\dagger \kett{ \qnums } &= \begin{cases}
t \geq 0 : & \kett{t+1, \cdots }, \\
t < 0 : & \kett{ t+1,\ell_{12}+1, \ell_{23} + 1, \ell_{31} + 1, \cdots } + \kett{ t+1,\ell_{21}+1, \ell_{32} + 1 , \ell_{13} + 1, \cdots  },
\end{cases} \label{eq: TAdagger action} \\
T_B^\dagger \kett{ \qnums } &= \begin{cases}
t \leq 0 : & \kett{t-1, \cdots }, \\
t > 0 : & \kett{ t-1,\ell_{12}+1, \ell_{23} + 1, \ell_{31} + 1, \cdots } + \kett{ t-1, \ell_{21}+1, \ell_{32} + 1 , \ell_{13} + 1, \cdots } .
\end{cases}\label{eq: TBdagger action}
\end{align}
$\theta$ in the formulas above denotes the Heaviside function.

The most relevant, nontrivial operators consist of the corner operators that lower at least one of the $P_i$ or $Q_i$, that is, all the corner operators except for $L_{ij}^\dagger$ pure-creation operators.
The matrix representations of these corner operators are as follows.
\begin{align}
    J^\dagger_{ij} \kett{\mathbf{q} } &= \begin{cases}
        t \geq 0: & \kett{t+1,l_{kj}-1, \cdots} \epsilon_{ijk} l_{kj} ,\\
        t < 0:    & \kett{t+1,l_{ji}+1,l_{ik}+1,\cdots} \epsilon_{ijk} (l_{kj} + |t|) 
    \\
    & + \kett{t+1,l_{ij}+1,l_{jk}+1,l_{ki}+1,l_{kj}-1,\cdots} \epsilon_{ijk} l_{kj} .\\
    \end{cases} \label{eq: Jij dagger result} \\
    K^\dagger_{ij} \kett{\mathbf{q} } &= \begin{cases}
        t \leq 0: & \kett{t-1,l_{jk}-1, \cdots} \epsilon_{ijk} l_{jk} ,\\
        t > 0:    & \kett{t-1,l_{ij}+1,l_{ki}+1,\cdots} \epsilon_{ijk} (l_{jk} + |t|) \\
    & + \kett{t-1,l_{jk}-1,l_{ji}+1,l_{kj}+1,l_{ik}+1,\cdots} \epsilon_{ijk} l_{jk} .\\
    \end{cases} \label{eq: Kij dagger result} \\
    \label{eq: Nij result}
    N_{ij} \kett{ \mathbf{q} } &= \kett{l_{jk}-1,l_{ik}+1,\cdots} \Bigl( \tfrac{l_{jk} (l_{ji}+\sigma_{jk}+|t|+1)}{\sigma_{ij}+\sigma_{jk}+|t| + 1} \Bigr) \nonumber \\
    & \quad + \kett{l_{ij}+1,l_{ki}+1,l_{ji}-1,l_{kj}-1,\cdots} \Bigl( \tfrac{- l_{ji} l_{kj}}{\sigma_{ij}+\sigma_{jk}+|t| + 1} \Bigr) .\\
    \label{eq: Mij result}
    M_{ij} \kett{ \mathbf{q} } &= \kett{l_{kj}-1,l_{ki}+1,\cdots} \Bigl( \tfrac{l_{kj} (l_{ij}+\sigma_{jk}+|t|+1)}{\sigma_{ij}+\sigma_{jk}+|t| + 1} \Bigr) \nonumber \\
    & \quad + \kett{l_{ji}+1,l_{ik}+1,l_{ij}-1,l_{jk}-1,\cdots} \Bigl( \tfrac{- l_{ij} l_{jk}}{\sigma_{ij}+\sigma_{jk}+|t| + 1} \Bigr) .\\
    \label{eq: Lij result}
    L_{ij} \kett{ \mathbf{q} } &=
     \kett{l_{ij}-1,\cdots} \tfrac{l_{ij}}{1+|t|+\sigma_{ij}+\sigma_{ik}} \Bigl\{ (2+|t|+l_{ij} + l_{kj} + \sigma_{ik} )(1+|t|+l_{ik}+\sigma_{ij}) \nonumber \\
    & \qquad \qquad \qquad \qquad \qquad \qquad \quad - \tfrac{(1+l_{ik}) l_{jk} (1+|t|+l_{ij}+\sigma_{ik})
    +l_{ji}(1+l_{ki})(1+|t|+l_{jk}+\sigma_{ij}) 
    }{1+|t|+\sigma_{ij}+\sigma_{jk}} \Bigr\} \nonumber \\
    & \quad + \kett{l_{ij}-2,l_{jk}-1,l_{ki}-1,l_{ji}+1,l_{kj}+1,l_{ik}+1,t} \tfrac{(l_{ij}-1)l_{ij} l_{jk} l_{ki}}{(1+|t|+\sigma_{ij}+\sigma_{ik})(1+|t|+\sigma_{ij}+\sigma_{jk})} \nonumber \\
    & \quad + \kett{l_{ij},l_{jk}+1,l_{ki}+1,l_{ji}-1,l_{kj}-1,l_{ik}-1,t} \Bigl(\tfrac{-l_{ik} l_{ji} l_{kj}}{1+|t|+\sigma_{ij}+\sigma_{ik}}\Bigr) \Bigl( 1+ \tfrac{2+|t|+l_{ij} + \sigma_{ik}}{1+|t|+\sigma_{ij} + \sigma_{jk}} \Bigr) . \\
    \label{eq: Jij result}
    J_{ij} \kett{ \mathbf{q} } &= \begin{cases}
        t \leq 0: &  
    \kett{t-1,l_{ji}-1,l_{ik}-1,\cdots} \epsilon_{ijk} 
    \Bigl( \tfrac{l_{ji} l_{ik}(2+|t|+\sigma_{ij}+\sigma_{ik}+l_{jk})}{1+|t|+\sigma_{ij}+\sigma_{ik}} \Bigr) \\
    & + \kett{t-1,l_{ij}-1,l_{jk}-1,l_{ki}-1,l_{kj}+1,\cdots} \epsilon_{ijk} 
    \Bigl( \tfrac{l_{ki} l_{ij} l_{jk}}{1+|t|+\sigma_{ij}+\sigma_{ik}} \Bigr) , \\
    t > 0:
    & \kett{t-1,l_{ij}+1,l_{jk}+1,l_{ki}+1,l_{ji}-1,l_{kj},l_{ik}-1} \epsilon_{ijk} \Bigl( \tfrac{l_{ji} l_{ik} (2+2|t| +\sigma_{ij}+\sigma_{ik}+l_{jk})}{1+|t|+\sigma_{ij}+\sigma_{ik}} \Bigr) \\
    & +\kett{t-1,l_{ij}-1,l_{jk}-1,l_{ki}-1,l_{ji}+1,l_{kj}+2,l_{ik}+1} \epsilon_{ijk} \Bigl( \tfrac{l_{ki} l_{ij} l_{jk}}{1+|t|+\sigma_{ij}+\sigma_{ik}} \Bigr) \\
    & + \kett{t-1,l_{kj}+1,\cdots} \epsilon_{ijk} \Bigl\{ l_{ji} l_{ik} + |t| (1+|t|+l_{ji}+l_{ik}+l_{jk}) \\
    & \qquad \qquad \qquad \qquad \qquad \qquad +\tfrac{l_{ji} l_{ik} (1+l_{jk}) + l_{ki} l_{ij} (|t|+l_{jk})}{1+|t|+\sigma_{ij}+\sigma_{ik}} \Bigr\} . \\
    \end{cases} \\
    \label{eq: Kij result}
    K_{ij} \kett{ \mathbf{q} } &= \begin{cases}
        t \geq 0: &  
    \kett{t+1,l_{ij}-1,l_{ki}-1,\cdots} \epsilon_{ijk} 
    \Bigl( \tfrac{l_{ij} l_{ki}(2+|t|+\sigma_{ij}+\sigma_{ik}+l_{kj})}{1+|t|+\sigma_{ij}+\sigma_{ik}} \Bigr) \\
    & + \kett{t+1,l_{jk}+1,l_{ji}-1,l_{kj}-1,l_{ik}-1,\cdots} \epsilon_{ijk} 
    \Bigl( \tfrac{l_{ik} l_{ji} l_{kj}}{1+|t|+\sigma_{ij}+\sigma_{ik}} \Bigr) , \\
    t < 0:
    & \kett{t+1,l_{ij}-1,l_{jk},l_{ki}-1,l_{ji}+1,l_{kj}+1,l_{ik}+1} \epsilon_{ijk} \Bigl( \tfrac{l_{ij} l_{ki} (2+2|t| +\sigma_{ij}+\sigma_{ik}+l_{kj})}{1+|t|+\sigma_{ij}+\sigma_{ik}} \Bigr) \\
    & + \kett{t+1,l_{ij}+1,l_{jk}+2,l_{ki}+1,l_{ji}-1,l_{kj}-1,l_{ik}-1} \epsilon_{ijk} \Bigl( \tfrac{l_{ik} l_{ji} l_{kj}}{1+|t|+\sigma_{ij}+\sigma_{ik}} \Bigr) \\
    & + \kett{t+1,l_{jk}+1,\cdots} \epsilon_{ijk} \Bigl\{ l_{ij} l_{ki} + |t| (1+|t|+l_{ij}+l_{ki}+l_{kj}) \\
    & \qquad \qquad \qquad \qquad \qquad \qquad +\tfrac{l_{ij} l_{ki} (1+l_{kj}) + l_{ik} l_{ji} (|t|+l_{kj})}{1+|t|+\sigma_{ij}+\sigma_{ik}} \Bigr\} .\\
    \end{cases}
\end{align}

Lastly, there are the operators that involve all three vertex legs, such as $T_A$ and $T_B$.
Their matrix representations are not written out here explicitly because they can be derived from commutator identities, see \eqref{eq: TA and TB operator identities}, \eqref{eq: ++- corner operator identities}, and \eqref{eq: +-- corner operator identities}.

When working with the ISB multiplets, one finds that there are yet more ways to contract them into gauge-invariant operators associated with the vertex, but they can always be related to gauge-singlet operators that have been named above. (An example of this will be given in Eq.~\eqref{eq: out of normal order example} later.) Thus, the operators covered above are the essential ones to have analytic control over.

\subsection{Programming implementation}
\label{sec: resultscode}
For illustrative purposes, a companion Mathematica notebook has been prepared that implements all of the above operators, provided as Supplemental Material \cite{SM}.
The program follows a similar structure to the companion code provided for Part I, but instead of working in the underlying Hilbert space of Schwinger bosons---the tensor product of 18 harmonic oscillators---it works in the space of seven LSH quantum numbers.
The gauge-singlet operators are implemented as functions whose inputs and outputs are Hilbert-space kets.
In alignment with the presentation in Secs.~\ref{sec: commutators} and \ref{sec: resultsanalytic}, some operators are hard-coded with formulas while others are implemented by taking advantage of commutator identities.

With the ability to do faster calculations in terms of strictly LSH degrees of freedom, example commands are included for tasks such as state normalization, Gram-Schmidt orthogonalization, and constructing the eigenstates of the seventh Casimir operator ($C_T = T_A^\dagger T_B^\dagger T_A T_B$) that was discussed in Part I.

\section{\label{sec: methods}Methods and derivations}
 \label{sec: derivationspart2}
To derive the results in Sections \ref{sec: commutators} and \ref{sec: resultsanalytic}, calculations are done using the ISBs.
These have already been described in detail in Part 1, so we only review some essential details for completeness.
The gauge-singlet operators are then expressed in terms of the ISBs.
The algebraic identities in Sec.~\ref{sec: commutators} follow from using the elementary properties of ISBs, and as noted earlier, we have omitted their proofs due to the methodological overlap with prior works.
The Sec.~\ref{sec: resultsanalytic} results follow from more substantial calculations that will be explained below, although the derivations are expedited with the algebra of Sec.~\ref{sec: commutators} at hand.

\subsection{A brief reminder of SU(3) irreducible Schwinger bosons \label{sec:isb}}
\noindent

With the ISBs being the underlying foundation for the SU(3) LSH formulation, we will below give a concise review of the most essential properties needed to derive all subsequent results.
The reader is referred to Part I and Refs.~\cite{Mathur:2004kr, Anishetty:2009ai, Anishetty:2009nh} for more details.

For each leg $i$, there is an associated pair of triplets of ISB operators $A(i)^\alpha$ and $B(i)_\alpha$ for $\alpha=1,2,3$ (a total of six modes per leg, or twelve modes per link when considering a lattice of connected vertices).
The color index $\alpha$ is said to be a fundamental (antifundamental) index if it is used as a lower (upper) index.
When only one vertex leg is being discussed, we often suppress the $i$ label: $A(i)^\alpha \to A^\alpha$ and $B(i)_\alpha \to B_\alpha$.
The $A$ and $B$ triplets are themselves constructed from triplets of ordinary simple harmonic oscillator modes; however, the derivations below do not have to invoke these relationships.
The only properties needed are the effective identities 
\begin{subequations}
\begin{align}
    A^\dagger_\alpha A^\alpha &= P,\\
    B^{\dagger \, \alpha} B_\alpha &= Q,\\
    A^\dagger_\alpha B^{\dagger \, \alpha} &= 0,  \\
     A^\alpha B_\alpha &= 0,
\end{align}
\end{subequations}
along with the effective commutators
\begin{subequations}
\begin{align}
    [ A^\dagger_\alpha, A^\dagger_{\beta}]=[ B^{\dagger\alpha}, B^{\dagger\beta}]=[A^\dagger_\alpha, B^{\dagger\beta} ]&=0, \\
    [ A^\alpha, A^{\beta}]=[B_\alpha,B_\beta]=[A^\alpha,B_\beta]&= 0 ,\\
    [A^\alpha, A^\dagger_\beta] &= \left( \delta^\alpha_{\beta}-\frac{1}{P+Q+2}B^{\dagger\alpha}B_\beta \right), \\
    [B_\alpha, B^{\dagger\beta}] &=  \left( \delta_\alpha^{\beta}-\frac{1}{P+Q+2}A^{\dagger}_{\alpha}A^\beta \right), \\
    [A^\alpha,B^{\dagger\beta}] &=   -\frac{1}{P+Q+2}B^{\dagger\alpha}A^\beta, \\
    [B_\alpha,A^\dagger_\beta] &=   -\frac{1}{P+Q+2}A^{\dagger}_\alpha B_\beta. 
\end{align}
\end{subequations}
Because the quantity $(P + Q + 2)^{-1}$ is so frequently occurring, we also introduced the $F_i$ operators in \eqref{eq: Fi definition}.
Note that ISB operators belonging to different vertex legs always identically commute.
There exists a unique bosonic vacuum state for the vertex, $\ket{0}$, which is characterized by
\begin{align}
    A(i)^\alpha \ket{0} = B(i)_\alpha \ket{0} &= 0 \quad \text{for all $i$ and $\alpha$.}
\end{align}
The state $\ket{0}$ is used as a reference state from which all gauge-invariant states can be constructed by suitable applications of gauge-invariant creation operators.

\subsection{Key operators constructed from ISBs}
\noindent

SU(3)-invariant operators are formed by contracting indices of the SU(3) ISB triplets, the $\delta$ tensor, and/or the rank-three $\epsilon$ tensor.
The simplest operators are classified as bilinears or trilinears according to whether the singlet is formed by contracting two triplets or three triplets into overall SU(3) singlet objects.

The simplest of these contractions are the number operators,
\begin{align}
    P_i &= A(i)^\dagger_\alpha A(i)^\alpha , \quad    Q_i = B(i)^{\dagger \, \alpha} B(i)_\alpha .
\end{align}
To construct the vertex Hilbert space, one has the purely creation-type bilinears
\begin{align}
    L_{ij}^\dagger &\equiv A(i)^\dagger _\alpha B(j)^{\dagger \alpha}. \\
\end{align}
Then there are the mixed-type bilinears ($i \neq j$)
\begin{align}
    N_{ij} &\equiv A(i)^{\dagger}_{\alpha} A(j)^{\alpha} , \quad   M_{ij} \equiv B(i)^{\dagger \, \alpha} B(j)_{\alpha} .
\end{align}

For the trilinear contractions, the simplest are the purely creation-type ones that are also used to construct the gauge invariant Hilbert space:
\begin{subequations}
\begin{align}
    T_A^\dagger &\equiv \epsilon^{\alpha \beta \gamma} A(1)^\dagger_\alpha A(2)^\dagger_\beta A(3)^\dagger_\gamma ,\\
    T_B^\dagger &\equiv \epsilon_{\alpha \beta \gamma} B(1)^{\dagger \, \alpha} B(2)^{\dagger \, \beta} B(3)^{\dagger \, \gamma} .
\end{align}
\end{subequations}
For brevity, when dealing with $\epsilon$-contractions of ISBs, we often use a notation that suppresses the contracted color indices and puts the leg labels as subscripts.
Hence, we can write
\begin{align}
    T_A^\dagger &= \epsilon A_1^\dagger A_2^\dagger A_3^\dagger , \quad 
    T_B^\dagger = \epsilon B_1^\dagger B_2^\dagger B_3^\dagger .
\end{align}
The associated trilinear annihilation operators are 
\begin{subequations}
\begin{align}
    T_A &\equiv \epsilon_{\alpha \beta \gamma} A(1)^\alpha A(2)^\beta A(3)^\gamma = \epsilon A_1 A_2 A_3 , \\
    T_B &\equiv \epsilon^{\alpha \beta \gamma} B(1)_{\alpha} B(2)_{\beta} B(3)_{\gamma} = \epsilon B_1 B_2 B_3 .
\end{align}
\end{subequations}
As the $L_{ij}$ and $T_{A/B}$ operators obey the effective ``Mandelstam identity,'' that is,
\begin{align}
    T_A^\dagger T_B^\dagger &= L^\dagger_{12}L^\dagger_{23}L^\dagger_{31}+L^\dagger_{21}L^\dagger_{32}L^\dagger_{13} , \label{eq: SU(3) Mandelstam}
\end{align}
it follows that the simultaneous excitation of both trilinear modes can always be decomposed into linear combinations of bilinear flux excitations.

There are also a number of trilinear gauge-singlet operators that have mixed creation-annihilation actions.
Limiting ourselves to normal-ordered corner operators, we have
\begin{subequations}
\begin{alignat}{4}
    J_{ij}^\dagger &= \epsilon A_i^\dagger A_j^\dagger B_j & &\equiv \epsilon^{\alpha \beta \gamma} A(i)^\dagger_\alpha A(j)^\dagger_\beta B(j)_\gamma , \\
    K_{ij}^\dagger &= \epsilon B_i^\dagger B_j^\dagger A_j & &\equiv \epsilon_{\alpha \beta \gamma} B(i)^{\dagger \, \alpha} B(j)^{\dagger \, \beta} A(j)^{\gamma} , \\
    J_{ij} &= - \epsilon B_j^\dagger A_j A_i & &\equiv - \epsilon_{\alpha \beta \gamma} B(j)^{\dagger \, \alpha} A(j)^\beta A(i)^\gamma , \\
    K_{ij} &= - \epsilon A_j^\dagger B_j B_i & &\equiv - \epsilon^{\alpha \beta \gamma} A(j)^{\dagger}_{\alpha} B(j)_\beta B(i)_\gamma .
\end{alignat}
\end{subequations}
Corner operators that are not normal-ordered can be simply related to those above. For example,
\begin{align}
    \epsilon A_i^\dagger B_j A_j ^\dagger &\equiv \epsilon^{\alpha \beta \gamma} A(i)^\dagger_\alpha B(j)_\beta A(j)^\dagger_\gamma \nonumber \\
    &= - \left( \frac{P_j + Q_j + 3}{P_j + Q_j + 2} \right) \epsilon A_i^\dagger A_j^\dagger B_j . \label{eq: out of normal order example}
\end{align}
There are also mixed-type singlet contractions involving all three vertex legs,
such as $\epsilon^{\alpha \beta \gamma} A(i)^\dagger_\alpha A(j)^\dagger_\beta B(k)_\gamma$ which was abbreviated as $\epsilon A_i^\dagger A_j^\dagger B_k$ in \eqref{eq: ++- corner operator identities}.
It has already been noted earlier how these operators play a relatively minor role, as they are derivable from corner operators.

\subsection{Definition of basis}
To prove the formulas for SU(3) singlet operators applied to LSH basis kets, it is necessary to recall how that basis is actually defined.
The naive LSH states are defined as
\begin{align}
    \label{eq:naive basis kets}
    \kett{\ell_{12}, \ell_{23}, \ell_{31}, \ell_{21}, \ell_{32}, \ell_{13}, t } &\equiv L_{12}^{\dagger \, \ell_{12}} L_{23}^{\dagger \, \ell_{23}} L_{31}^{\dagger \, \ell_{31}} L_{21}^{\dagger \, \ell_{21}} L_{32}^{\dagger \, \ell_{32}} L_{13}^{\dagger \, \ell_{13}} \times \begin{cases}
        T_A^{\dagger \,  t} \ket{0}, & t \geq 0 \\ 
        T_B^{\dagger \, -t} \ket{0}, & t < 0
    \end{cases} \\
    \ell_{ij} &\in \{ 0, \ 1, \ 2, \ 3, \, \ldots \} , \\
    t &\in \{ 0, \ \pm 1, \ \pm 2 , \ \pm 3, \ \ldots \}
\end{align}
\footnote{In this work, any operator raised to the zeroth power is understood to be shorthand for the identity operator: $(L_{IJ}^\dagger)^0 = (T_A^\dagger)^0 = (T_B^\dagger)^0 \equiv 1$.}.
As mentioned earlier, the notation $\kett{*}$ is used to indicate a basis ket that is not necessarily normalized to unity.
In this notation, the total vacuum ket $\ket{0}$ may also be written as
\begin{align}
    \kett{0} &\equiv \ket{0} = \kett{0 , 0 , 0 , 0 , 0 , 0 , 0 }.
\end{align}
The vacuum is normalized to unity by definition: $\bbrakett{0|0}=\braket{0|0}=1$.

The above construction is a direct generalization of the SU(2) trivalent vertex Hilbert space \cite{Raychowdhury:2019iki}.
In Part I, we have discussed the fact that for SU(3), in contrast to SU(2), the basis states simply constructed from products of pure-creation singlets are not all orthogonal to each other.
The LSH ``quantum numbers'' therefore fail to be ideal quantum numbers for SU(3), in the sense that states with different quantum numbers are not necessarily orthogonal. It is for this reason we refer to the above basis as ``naive.''
Nonetheless, we argued in Part I that the naive basis is still a valid one, i.e., the states are linearly independent and complete, but not overcomplete.
Our argument was not a mathematical proof, but the results and proofs reported in the present work demonstrate completeness for Hamiltonian dynamics;
the basis in Eq.~\eqref{eq:naive basis kets} evidently spans the dynamical Hilbert space because it is closed under the application of any SU(3) singlet operator that could appear in the Hamiltonian.

\subsection{Proofs of the matrix-representation formulas}
The proofs of \eqref{eq: Lijdagger action}, \eqref{eq: TAdagger action}, and \eqref{eq: TBdagger action}, which involve pure-creation operators exclusively, are relatively trivial. They follow immediately from the definitions \eqref{eq:naive basis kets} along with the identities \eqref{eq: pure-creation commutators} and \eqref{eq: SU(3) Mandelstam}. Equations \eqref{eq: Jij dagger result}-\eqref{eq: Kij result} involve operators that all have some ISB-annihilation action associated with them, and this makes their proofs nontrivial.

The strategy we adopt to compute $\mathcal{O} \kett{\mathbf{q} }$ for general $\mathbf{q}$ consists of the following algorithm.
First we will present the algorithm in general terms, then we will describe its application to the corner operators of interest.\\

Let $q_r$ for $r=1,2, ...$,  denote an ordering of the quantum numbers in $\mathbf{q}$;
the choice of ordering is a separate consideration to be discussed further below.
Starting from the total bosonic vacuum $\ket{0}$, the degrees of freedom $q_r$ are incorporated one by one into the state acted on by $\mathcal{O}$, until finally $\mathcal{O} \kett{ q_1, q_2, \ldots }$ is known for any valid choice of the quantum numbers $\mathbf{q}$.
This occurs through a sequence of iterations, one for each value of $r$.
In the $r^{\mathrm{th}}$ iteration, it is assumed that $\mathcal{O} \kett{q_{1},q_{2},...,q_{r-1}}$ is known, and the objective is to include the next quantum number, $q_r$ by computing $\mathcal{O}\kett{q_1,q_{2}, \ldots , q_{r-1},q_r}$.
For brevity, let $\ket{\psi_r} = \kett{q_1,q_2,\ldots , q_r}$  denote the state containing the first $r$ quantum numbers.
The state $\ket{\psi_0}$ is identified with the total vacuum state $\ket{0}$.

In the $r^{\mathrm{th}}$ iteration:
\begin{itemize}
    \item $\mathcal{O} \ket{\psi_{r-1}}$ is assumed to be known. Let $\mathcal{B}^\dagger_r$ denote the creation operator associated with quantum number $q_r$. Note that if $q_r=t$, then $B_r^\dagger$ refers to $T_A^\dagger$ ($T_B^\dagger$) for $t>0$ ($t<0$).
    \item The evaluation of $\mathcal{O} \ket{\psi_{r}}$ will proceed based on the mathematical form (expression) of $[\mathcal{O},\mathcal{B}_r^\dagger ]$. 
    Here is where one should refer to the commutators provided in \ref{sec: commutators}.
    We identify two relevant possibilities for the form of $ [ \mathcal{O}, \mathcal{B}^\dagger_r ]$:
    \begin{enumerate}
        \item [F1)] $[ \mathcal{O}, \mathcal{B}^\dagger_r ] $ can be written as $ \mathcal{D} \mathcal{B}^\dagger_r \mathcal{O}$ where $\mathcal{D}$ is some function of $P_i$ and $Q_i$ occupation numbers only. This form includes the possibility $[ \mathcal{O},\mathcal{B}^\dagger_r]=0$.
        \item [F2)] $[ \mathcal{O}, \mathcal{B}^\dagger_r ] $ is a function of operators such that $[ \mathcal{O}, \mathcal{B}^\dagger_r ] \ket{\psi_r} $ can be evaluated using prior results, or can otherwise be computed without the knowledge of $\mathcal{O} \ket{\psi_r}$.
    \end{enumerate}
    \item In the case of form F1, the commutator can be arranged to say
        $\mathcal{O} \mathcal{B}^\dagger_r = (1 + \mathcal{D}) \mathcal{B}^\dagger_r \mathcal{O}$,
    which makes it possible to commute $\mathcal{O}$ through the powers of $\mathcal{B}^\dagger_r$:
    \begin{align}
        \mathcal{O} \ket{\psi_r} &= \mathcal{O} \mathcal{B}^\dagger_r {}^{|q_r|} \ket{\psi_{r-1}} \nonumber \\
        &= ((1+\mathcal{D})B^\dagger_r )^{|q_r|} \mathcal{O} \ket{\psi_{r-1}}. \label{eq: form 1}
    \end{align}
    Since $\mathcal{O} \ket{\psi_{r-1}}$ is known, as are the matrix elements for any power of $\mathcal{B}^\dagger_r$ (Eqs.~\eqref{eq: Lijdagger action}-\eqref{eq: TBdagger action}), all that remains is to evaluate the diagonal-operator factors of $1+\mathcal{D}$.
    \item In the case of form F2, one uses the commutator identity $[ O_1, O_2^p]=\sum_{n=0}^{p-1} O_2 ^{p-1-n} [ O_1 , O_2 ] O_2^n$ to write
    \begin{align}
        \mathcal{O} \ket{\psi_r} & = \mathcal{O} \mathcal{B}^\dagger_r {}^{|q_r|} \ket{\psi_{r-1}} \nonumber \\
        &= \mathcal{B}^\dagger_r {}^{|q_r|} \mathcal{O} \ket{\psi_{r-1}} + 
        \sum_{n=0}^{|q_r|-1} \mathcal{B}^\dagger_r {}^{|q_r|-1-n} [ \mathcal{O} , \mathcal{B}^\dagger_r ] \mathcal{B}^\dagger_r {}^{n} \ket{\psi_{r-1}} . \label{eq: form 2}
    \end{align}
    The first term, $\mathcal{B}^\dagger_r {}^{|q_r|} \mathcal{O} \ket{\psi_{r-1}}$, can be evaluated since $\mathcal{O} \ket{\psi_{r-1}}$ is known and the matrix elements for any power of $\mathcal{B}^\dagger_r$ are also known.
    Regarding the second term, the $\sum_n$ can be evaluated because
    (i) by assumption, $[ \mathcal{O}, \mathcal{B}^\dagger_r ] \mathcal{B}^\dagger_r {}^{n} \ket{\psi_{r-1}} $ is known/implied by prior results or it can otherwise be computed without the knowledge of $\mathcal{O} \ket{\psi_r}$; and
    (ii) the subsequent application of $\mathcal{B}^\dagger_r {}^{|q_r|-1-n}$ only requires the established matrix elements of pure-creation operators.\footnote{The summation over $n$ frequently ends up requiring the identity $\sum_{n=0}^{\ell-1} \frac{1}{(1+m+n)(2+m+n)} = \frac{\ell}{(1+m)(1+m+\ell)}$.}
\end{itemize}
These iterations continue until all desired quantum numbers have been included in the state acted on by $\mathcal{O}$.\\

It was remarked that the order in which the quantum numbers are incorporated is an important consideration.
This is related to the stipulation associated with form F2 -- that $[ \mathcal{O}, \mathcal{B}^\dagger_r ] \ket{\psi_r} $ can be computed \emph{without knowledge of $\mathcal{O} \ket{\psi_r}$}.
A poor choice of ordering for $\mathbf{q}$ may result in circular logic in which the evaluation of $\mathcal{O} \ket{\psi_r}$ cannot proceed without $\mathcal{O} B_r^\dagger {}^n \ket{\psi_{r-1}}$.
It is most convenient when $[ \mathcal{O}, \mathcal{B}^\dagger_r ] \ket{\psi_r} $ can be evaluated using prior results, but we do not think it is always possible to arrange for this -- sometimes, a side calculation is needed, and the ordering will also influence how laborious such side calculations are.
We often found it convenient to start by first including the quantum numbers whose associated creation operators commute with $\mathcal{O}$.

The algorithm will now be illustrated for the main corner operators of interest.

\subsubsection{Proof for $J_{ij}^\dagger$ and $K_{ij}^\dagger$}
We first demonstrate the calculational algorithm for the operators $\mathcal{O} =J_{ij}^\dagger$.
\begin{itemize}
    \item Before starting, note that $\mathcal{O} \ket{0}=J_{ij}^\dagger\ket{0}=0$ due to the annihilation content $B(j)$ in $J^\dagger_{ij}$.
    \item Consider the ordering
    $(q_1,q_2,q_3,q_4,q_5,q_6,q_7)\equiv
    (l_{ik},l_{ji},l_{ki},l_{jk},l_{ij},t,l_{kj})$.
    \item Iteration 1: $q_r=l_{ik}$. The commutator $[ \mathcal{O}, B_r^\dagger] = [ J^\dagger_{ij} , L^\dagger_{ik}] = 0$, which matches form F1 with $\mathcal{D}=0$. One finds $\mathcal{O} \ket{\psi_1} = J_{ij}^\dagger \kett{l_{ik}}= 0$.
    \item Iterations 2, 3, 4, 5: One can add in the quantum numbers $q_2=l_{ji}$, $q_3=l_{ki}$, $q_4=l_{jk}$,  and $q_5=l_{ij}$, each time using the same argument as Iteration 1, to eventually arrive at $\mathcal{O} \ket{\psi_5}=J^\dagger_{ij} \kett{l_{ik},l_{ji},l_{ki},l_{jk},l_{ij}} = 0$.
    \item Iteration 6: $q_r=t$. For this particular quantum number, one must proceed according to the sign of $t$:
    \begin{enumerate}
        \item $t>0:$ In this scenario, $[ \mathcal{O}, B^\dagger_r] = [ J^\dagger_{ij} , T_A^\dagger ] = 0$, which leads to $J^\dagger_{ij} \kett{q_1,\ldots,q_5,t>0} = 0$.
        \item $t<0:$ In this scenario, $[ \mathcal{O}, B^\dagger_r] = [ J^\dagger_{ij} , T_B^\dagger ] =  \epsilon_{ijk} L^\dagger_{ik} L^\dagger_{ji}$, which corresponds to form F2. Evaluating the right-hand side of \eqref{eq: form 2} is trivial, and leads to $J^\dagger_{ij} \kett{q_1,\ldots,q_5,t<0}= \epsilon_{ijk} |t| \kett{l_{ik}+1,
        l_{ji}+1,
        l_{ki},
        l_{jk},
        l_{ij},
        t+1}$.
    \end{enumerate}
    Putting all cases of $t$ together, one has
    \begin{align}
        \label{eq: JDagger round 6}
        \mathcal{O} \ket{\psi_6} &=J_{ij}^\dagger \kett{ l_{ik},l_{ji},l_{ki},l_{jk},l_{ij},t } =
        \begin{cases}
            t\geq 0: & 0,\\
            t< 0: & \epsilon_{ijk} |t| 
            \kett{ l_{ik}+1,l_{ji}+1,l_{ki},l_{jk},l_{ij},t+1 }.
        \end{cases}
    \end{align}
    \item Iteration 7: $q_r=l_{kj}$. In this case, $[\mathcal{O},\mathcal{B}^\dagger_r]=[J^\dagger_{ij},L^\dagger_{kj}]=\epsilon_{ijk} T^\dagger_A$, which matches form F2.
    The application of $T^\dagger_A$ is given by \eqref{eq: TAdagger action}, and after evaluating the $\sum_n$, one arrives at the result reported in \eqref{eq: Jij dagger result}.
\end{itemize}
The formula for $K^\dagger_{ij}$ in \eqref{eq: Kij dagger result} is obtained by using charge-conjugation symmetry, $A\leftrightarrow B$ for all the Schwinger-boson modes.

\subsubsection{Proof for $N_{ij}$ and $M_{ij}$}
Equipped with the formulas for $J_{ij}^\dagger$ and $K_{ij}^\dagger$, one can derive \eqref{eq: Nij result} and \eqref{eq: Mij result}. Consider $\mathcal{O}=N_{ij}$.
\begin{itemize}
    \item Before starting, note that $\mathcal{O} \ket{0}=N_{ij}\ket{0}=0$ due to the annihilation content $A(j)$ in $N_{ij}$.
    \item Consider the ordering
    $(q_1,q_2,q_3,q_4,q_5,q_6,q_7)\equiv
    (l_{ik},l_{ki},t,l_{jk},l_{ji},l_{kj},l_{ij})$.
    \item Iterations 1, 2: For quantum numbers $q_1=l_{ik}$ and $q_2=l_{ki}$, one can use the vanishing commutators in \eqref{eq: N and M Commutators combined} that match form F1 to conclude that $ \mathcal{O}\ket{\psi_2}=N_{ij} \kett{l_{ik},l_{ki}}=0$.
    \item Iteration 3: $q_r=t$. For $t>0$ ($t<0$), the relevant commutator is $[N_{ij}, T^\dagger_{A}]$ ( $[N_{ij},T^\dagger_B]$). For either case of $t$, $[\mathcal{O},B_r^\dagger]$ as it is expressed in \eqref{eq: N and M Commutators combined} matches the form F2, which apparently calls for matrix elements that have not yet been derived.
    However, it is easy to see that \[\left[N_{ij}, \mathcal{B}^\dagger_r\right] \mathcal{B}_r^\dagger {}^{n}\kett{l_{ik},l_{ki}} = 0\] for $t>0$ ($t<0$) because of the $B(j)$ and $(A(j))$ operators on the far right of the commutator. One thus arrives at
        $\mathcal{O} \ket{\psi_3} = N_{ij}\kett{l_{ik},l_{ki},t}=0$.
    \item Iteration 4: $q_r=l_{jk}$.
    Considering $[N_{ij},L^\dagger_{jk}]$ from \eqref{eq: N and M Commutators combined}, the evaluation matches case F2 and requires knowing $M_{kj} \kett{l_{ik},l_{ki},t,l_{jk}}$, but this has not yet been derived. However, this quantity can be related to the matrix representation of $N_{ij}$ by symmetries:
    \begin{align*}
        M_{kj} \kett{l_{ik},l_{ki},t,l_{jk}} \ \overset{i \leftrightarrow k}{\longrightarrow} \ 
        M_{ij} \kett{l_{ki},l_{ik},t,l_{ji}} \overset{A\leftrightarrow B}{\longrightarrow}
        N_{ij} \kett{l_{ik},l_{ki},-t,l_{ij}}.
    \end{align*}
    The quantity $N_{ij} \kett{l_{ik},l_{ki},-t,l_{ij}}$ also has not yet been derived, but it can be obtained from the Iteration 3 result by temporarily treating $l_{ij}$ as the fourth quantum number to be included; the commutator $[N_{ij},L_{ij}^\dagger]$ matches form F1, so that one can deduce
    \begin{align}
        N_{ij} \kett{l_{ik},l_{ki},t,l_{ij}} &= 0  \\
        \Rightarrow M_{kj} \kett{l_{ik},l_{ki},t,l_{jk}} &= 0. \label{eq: Mkj on four}
    \end{align}
    Equipped with the above, the evaluation of \eqref{eq: form 2} can be carried out, with the final answer of
    \begin{align}
        \mathcal{O} \ket{\psi_4} = N_{ij}\kett{l_{ik},l_{ki},t,l_{jk}}=l_{jk}\kett{l_{ik}+1,l_{ki},t,l_{jk}-1}. \label{eq: Nij Mij round 4}
    \end{align}
    \item Iteration 5: $q_r=l_{ji}$.
    The commutator $[N_{ij},L^\dagger_{ji}]$ in \eqref{eq: N and M Commutators combined} matches case F2.
    Evaluating \eqref{eq: form 2} evidently requires knowing $M_{ij} \kett{l_{ik},l_{ki},t,l_{jk},l_{ji}}$, but this quantity has not yet been derived.
    It is not hard to show from the $M_{ij}$ commutators in \eqref{eq: N and M Commutators combined} that
    \begin{align}
        M_{ij} \kett{l_{ik},l_{ki},t,l_{jk},l_{ji} } &= L_{ki}^\dagger {}^{l_{ki}} L_{ik}^\dagger {}^{l_{ik}} ( (1-F_j)L_{ji}^\dagger) {}^{l_{ji}} M_{ij} \kett{t,l_{jk}}. \label{eq: Mij after commuting}
    \end{align}
    The quantity $M_{ij} \kett{t,l_{jk}}$ also has not yet been derived, but it turns out to be zero. To see this, one can apply our same algorithm with $\mathcal{O}=M_{ij}$, treating $t$ and $l_{jk}$ as the first and second quantum numbers to be included, respectively:
    \begin{enumerate}
        \item In the first iteration, note that $M_{ij}\kett{t}=0$ is implied by applying symmetries on \eqref{eq: Nij Mij round 4}.
        \item In the second iteration, the commutator $[M_{ij},L_{jk}^\dagger]$ has form F2, and evaluating the sum in \eqref{eq: form 2} requires knowing $M_{kj} \kett{t,l_{jk}}$. The earlier Eq.~\eqref{eq: Mkj on four} contains $M_{kj} \kett{t,l_{jk}}=0$ as a special case, so that one arrives at $M_{ij} \kett{t,l_{jk}}=0$.
    \end{enumerate}
    Equipped with $M_{ij} \kett{t,l_{jk}}=0$, Eq.~\eqref{eq: Mij after commuting} then indicates that 
        $M_{ij} \kett{l_{ik},l_{ki},t,l_{jk},l_{ji} } =0$.
    Finally, one has the necessary information to carry out the evaluation of \eqref{eq: form 2}, eventually arriving at 
    \begin{align}
        \mathcal{O} \ket{\psi_5} =N_{ij}\kett{l_{ik},l_{ki},t,l_{jk},l_{ji}}
        &=l_{jk}\kett{l_{ik}+1,l_{ki},t,l_{jk}-1,l_{ji}}. \label{eq: Nij Mij round 5}
    \end{align}
    \item Iteration 6: $q_r=l_{kj}$.
    The commutator $[N_{ij},L_{kj}^\dagger]$ matches form F2, and in the process of evaluating \eqref{eq: form 2}, one has to know
    $N_{kj} \kett{l_{ik},l_{ki},t,l_{jk},l_{ji},l_{kj}}$.
    The latter can be handled by first commuting $N_{kj}$ through the powers of $L_{kj}^\dagger$ and consolidating the diagonal operators, that is,
    \begin{align*}
        N_{kj} L_{kj}^\dagger {}^n &= ( (1-F_j) L_{kj}^\dagger )^n N_{kj} \\
        &= \left( \tfrac{1+P_j+Q_j}{2+P_j+Q_j} L_{kj}^\dagger \right)^n N_{kj} \\
        &= \left( \tfrac{1}{2+P_j+Q_j} L_{kj}^\dagger (2+P_j+Q_j) \right)^n N_{kj} \\
        &=  \tfrac{1}{2+P_j+Q_j} L_{kj}^\dagger {}^n (2+P_j+Q_j) N_{kj} \\
        &=  L_{kj}^\dagger {}^n \Bigl( \tfrac{2+P_j+Q_j}{2+n+P_j+Q_j}  \Bigr) N_{kj} . \\
        \Rightarrow N_{kj} 
        \kett{l_{ik},l_{ki},t,l_{jk},l_{ji},l_{kj}} &= L_{kj}^\dagger {}^{l_{kj}}\Bigl(\tfrac{2+P_j+Q_j}{2+n+P_j+Q_j} \Bigr) N_{kj} 
        \kett{l_{ik},l_{ki},t,l_{jk},l_{ji},l_{kj}} \\
        &= L_{kj}^\dagger {}^{l_{kj}}\Bigl(\tfrac{1+|t|+l_{ji}+l_{jk}}{1+n+|t|+l_{ji}+l_{jk}} \Bigr) N_{kj} \kett{
        l_{ik},l_{ki},t,l_{jk},l_{ji}} .
    \end{align*}
    Then, the quantity $N_{kj} \kett{l_{ik},l_{ki},t,l_{jk},l_{ji}}$ is obtained from \eqref{eq: Nij Mij round 5} by using $i\leftrightarrow k$.
    At that point, one has everything needed to evaluate \eqref{eq: form 2}, and will eventually arrive at
    \begin{align}
    N_{ij} \kett{l_{ik},l_{ki},t,l_{jk},l_{ji},l_{kj}} &= l_{jk} \kett{
    l_{ik}+1,l_{ki},t,l_{jk}-1,l_{ji},l_{kj}} \nonumber \\
    & \quad - \tfrac{l_{ji} l_{kj}}{(l_{ji}+l_{jk}+l_{kj}+|t|+1)} L^\dagger_{ij} \kett{l_{ik},l_{ki}+1,t,l_{jk},l_{ji}-1,l_{kj}-1} . \label{eq: Nij Mij round 6}
    \end{align}
    \item Iteration 7: $q_r=l_{ij}$.
    The commutator $[N_{ij},L_{ij}^\dagger]$ matches form F1.
    Using similar techniques as in Iteration 6, one can show
    \begin{align}
        N_{ij} L_{ij}^\dagger {}^{l_{ij}} \kett{l_{ik},l_{ki},t,l_{jk},l_{ji},l_{kj}} 
        &=  L^\dagger_{ij} {}^{l_{ij}} \left(\tfrac{l_{kj}+l_{ji}+l_{jk}+|t|+1}{l_{kj}+l_{ji}+l_{jk}+|t| + l_{ij} + 1}\right) N_{ij}\kett{
        l_{ik},l_{ki},t,l_{jk},l_{ji},l_{kj}} .
    \end{align}
    Then, inserting the Iteration 6 result, one arrives at the final answer reported in \eqref{eq: Nij result}.
\end{itemize}
The formula for $M_{ij}$ in \eqref{eq: Mij result} is obtained by using charge-conjugation symmetry, $A\leftrightarrow B$ for all the Schwinger-boson modes.

\subsubsection{Proof for $L_{ij}$}
To derive \eqref{eq: Lij result}, consider $\mathcal{O}=L_{ij}$. 
\begin{itemize}
    \item Before starting, note that $\mathcal{O}\ket{0}=L_{ij}\ket{0}=0$ due to the annihilation operators in $L_{ij}$.
    \item Consider the ordering
    $(q_1,q_2,q_3,q_4,q_5,q_6,q_7)\equiv
    (t,l_{ij},l_{ji},l_{kj},l_{jk},l_{ik},l_{ki})$.
    \item Iteration 1: $q_r=t$.
    In this iteration, no calculation is actually necessary; is easy to see that $L_{ij} \kett{t}=0$ is necessary  
    when $t> 0$ ($t< 0$) due to the annihilation components $B(j)$ ($A(i)$) in $L_{ij}$.
    \item Iteration 2: $q_r=l_{ij}$.
    The commutator $[L_{ij},L_{ij}^\dagger]$ from \eqref{eq: L Commutators combined} matches form F2.
    When applying $[L_{ij},L_{ij}^\dagger]$ to $\kett{t,l_{ij}}$, most of the operator terms can be evaluated using the established formulas for $P_i$, $Q_j$, $M_{ji}$, and $N_{ij}$.
    The only really lacking information is the quantity $L_{ji} \kett{t,l_{ij}}$.
    However, it is easy to argue that this must be zero because, for any choice of $l_{ij}$ and $t$, $L_{ji}=A(j)^\alpha B(i)_\alpha$ will contain annihilation operators that do not have corresponding creation operators to the right.
    One thus finds that $[L_{ij},L_{ij}^\dagger] \kett{t,l_{ij}} = (3 + 2 l_{ij} + |t|)\kett{t,l_{ij}}$.
    After carrying out the sum in \eqref{eq: form 2}, one arrives at
    \begin{align}
    L_{ij} \kett{t,l_{ij}}&= l_{ij} ( 2 + l_{ij} + |t| ) \kett{t,l_{ij}-1} . \label{eq: Lij round 2}
    \end{align}
    \item Iteration 3: $q_r=l_{ji}$.
    In this scenario, the commutator $[L_{ij},L_{ji}^\dagger]$, as given in \eqref{eq: L Commutators combined}, matches neither form F1 nor F2, but is instead a linear combination of the two.
    However, considering the established formulas for $N_{ji}$ and $M_{ij}$ and their application on the specific states $\kett{t,l_{ij},l_{ji}}$, one finds
    \begin{align}
        [L_{ij},L_{ji}^\dagger] \kett{t,l_{ij},l_{ji}} &= -F_i F_j L_{ji}^\dagger L_{ij} \kett{t,l_{ij},l_{ji}}.
    \end{align}
    Thus, the commutator virtually takes on the form F1, and the strategy is to commute $L_{ij}$ to the right through the powers of $L_{ji}^\dagger$. From the last equation, note
    \begin{align}
        L_{ij} L_{ji}^\dagger \kett{t,l_{ij},l_{ji}} &= (1 - F_i F_j) L_{ji}^\dagger L_{ij} \kett{t,l_{ij},l_{ji}} \\
        \Rightarrow L_{ij} (L_{ji}^\dagger)^n \kett{t,l_{ij},l_{ji}} &= ((1-F_i F_j)L_{ji}^\dagger)^n L_{ij} \kett{t,l_{ij},l_{ji}} \\
        \Rightarrow L_{ij} (L_{ji}^\dagger)^n \kett{t,l_{ij}} &= ((1-F_i F_j)L_{ji}^\dagger)^n L_{ij} \kett{t,l_{ij}} .
    \end{align}
    Setting $n=l_{ji}$ in the last line, one has
    \begin{align}
        \mathcal{O} \ket{\psi_3} = L_{ij} L_{ji}^\dagger {}^{l_{ji}} \kett{t,l_{ij}} &= ((1-F_i F_j)L_{ji}^\dagger)^{l_{ji}} L_{ij} \kett{t,l_{ij}} \nonumber \\
        &= ((1-F_i^2)L_{ji}^\dagger)^{l_{ji}} L_{ij} \kett{t,l_{ij}} . \label{eq: Lij commuted through powers of Lji dagger}
    \end{align}
    What remains to be evaluated at this point is the diagonal factors of $1-F_i^2$.
    It is helpful to observe that
    \begin{align}
        (1-F_i^2) L^\dagger_{ji} &= \Bigl( \tfrac{P_i + Q_i + 3}{P_i + Q_i + 2} \Bigr) L^\dagger_{ji} \Bigl( \tfrac{P_i + Q_i + 2}{P_i + Q_i + 3} \Bigr) , \\
        \Rightarrow  \bigl( (1-F_i^2) L^\dagger_{ji} \bigr)^n &= \Bigl( \tfrac{P_i + Q_i + 3}{P_i + Q_i + 2} \Bigr) L^\dagger_{ji} {}^n \Bigl( \tfrac{P_i + Q_i + 2}{P_i + Q_i + 3} \Bigr) \nonumber \\
        &= L_{ji}^\dagger {}^n \Bigl( \tfrac{P_i + Q_i + n + 3}{P_i + Q_i + n + 2} \Bigr) \Bigl( \tfrac{P_i + Q_i + 2}{P_i + Q_i + 3} \Bigr) . \label{eq: Lji power with simplified diagonal factors}
    \end{align}
    Using \eqref{eq: Lji power with simplified diagonal factors} in \eqref{eq: Lij commuted through powers of Lji dagger} and recalling the result in \eqref{eq: Lij round 2}, one arrives at
    \begin{align}
        L_{ij} \kett{t,l_{ij},l_{ji}}&= \tfrac{l_{ij}(l_{ij} + |t| + 1)(l_{ij} + l_{ji} + |t| + 2)}{(l_{ij} + l_{ji} + |t| + 1)}    \kett{t,l_{ij}-1,l_{ji}}. \label{eq: Lij round 3}
    \end{align}
    \item Iteration 4: $q_r=l_{kj}$.
    The commutator $[L_{ij},L_{kj}^\dagger]$ matches form F2, and applying it to states $\kett{t,l_{ij},l_{ji},l_{kj}}$ does not call for any unknown quantities.
    The end result of evaluating \eqref{eq: form 2} is
    \begin{align}
        L_{ij} \kett{t,l_{ij},l_{ji},l_{kj}} &= \tfrac{ l_{ij} (1+|t|+l_{ij}+l_{kj})(2+|t|+\sigma_{ij}+l_{kj})}{1+|t|+\sigma_{ij}+l_{kj}} 
        \kett{t,l_{ij}-1,l_{kj},l_{ji}} . \label{eq: Lij round 4}
    \end{align}
    \item Iterations 5, 6, 7: We next put in $l_{jk}$, followed by $l_{ik}$, and finally $l_{ki}$. Each round goes like Iteration 4, meaning the commutator $[\mathcal{O},\mathcal{B}^\dagger_r]$ matches form F2 and applying it to the state $ B^\dagger_r {}^{n} \ket{\psi_{r-1}}$ only requires matrix elements that have been established earlier. 
    The final result after Iteration 7 is that reported in \eqref{eq: Lij result}.
    If desired, the intermediate results for Iterations 5 and 6 may be obtained by setting appropriate variables equal to zero in \eqref{eq: Lij result}.
\end{itemize}

\subsubsection{Proof for $J_{ij}$ and $K_{ij}$}
To derive \eqref{eq: Jij result} and \eqref{eq: Kij result}, consider $\mathcal{O}=K_{ij}$. 
\begin{itemize}
    \item Before starting, note that $\mathcal{O}\ket{0}=K_{ij}\ket{0}=0$ due to the annihilation operators in $K_{ij}$.
    \item Consider the ordering
    $(q_1,q_2,q_3,q_4,q_5,q_6,q_7)\equiv
    (l_{jk},l_{kj},l_{ji},l_{ik},l_{ki},l_{ij},t)$.
    \item Iterations 1-7: All iterations can be carried out using the same techniques as in the previous proofs, and without the need for any side calculations.
    The only two iterations that deserve some mentioning are Iterations 2 and 7.
    In these iterations, the commutator $[\mathcal{O},B^\dagger_r]$ given in \eqref{eq: J and K commutators combined} involves a trilinear operator that has not been worked out explicitly, and one should use Eqs.~\eqref{eq: ++- corner operator identities} to express the trilinear operator in terms of corner operators whose formulas have already been established.
    The final result is as reported in \eqref{eq: Kij result}, and readers who wish to know the intermediate results after each iteration may set the appropriate variables equal to zero in that formula.
\end{itemize}
The formula for $J_{ij}$ is obtained from that of $K_{ij}$ by using charge-conjugation symmetry.

\section{Discussion\label{sec: discussionpart2}}

In this article, we have presented a subset of the commutation relations of LSH operators at a trivalent vertex and used it to derive a matrix representation of on-site gauge singlet operators in the SU(3) LSH framework. The results are obtained for a vertex equipped with the nonorthogonal SU(3) gauge-invariant basis that was introduced in Ref.~\cite{Kadam:2024ifg}.

In mathematical terms, the matrix representation of operators at the vertex means the following:
for any SU(3)-singlet operator $\mathcal{O}$ at the trivalent vertex and any basis ket $\kett{ \qnums }$, the state $\mathcal{O} \kett{ \qnums } $ can be expressed in the same basis:
\begin{align}
    \mathcal{O} \kett{ \qnums } &= \sum_{ \qnums' } [ \mathcal{O} ]_{ \qnums' , \qnums} \kett{ \qnums' }
\end{align}
for known coefficients $[ \mathcal{O} ]_{ \qnums' , \qnums}$ given in Sec.~\ref{sec: derivationspart2}.
This important task was also done in Refs.~\cite{Raychowdhury:2019iki,Kadam:2022ipf}, although it was much more straightforward for those theories, largely because there was no SU(3)-multiplicity problem to deal with.
This task is a prerequisite to computing the dynamics, whether that is through classical- or quantum-computing means.

In practice, to actually do the matrix arithmetic on computers one must in principle truncate the basis to finite dimensionality, and order it.
Recall that the usage of a basis that is not orthonormal can have unexpected consequences, for example, the matrix representation of a Hermitian operator not being equal to its conjugate-transpose. And at present it is not obvious how to circuitize a quantum simulation without a mapping between the computational basis states and orthogonal states of the lattice gauge theory.
Still, basis-independent quantities, such as the eigenvalues, will remain identical, and orthogonal bases can be constructed on a sector-by-sector basis using our provided code. Analytic work in this area can be tedious, as can be inferred from the derivations presented in the Sec.~\ref{sec: derivationspart2}, but calculations like those presented herein enable progress toward computing dynamics for the theory. The novelty of the current manuscript is to present the results in a ready-to-use format so that the community can work in terms of LSH variables and proceed further. The companion code facilitates the same further for practical implementations. 

It is important to note that the Clebsch-Gordon coefficients for SU(3) remain difficult to work with analytically, with their evaluation often being outsourced to software that will calculate them on a case-by-case basis \cite{Balaji:2025yua}. When working with 
Schwinger bosons, the ISB construction takes care of Clebsch-Gordon coefficients implicitly, however the authors have further experienced that the cost of numerically implementing the Schwinger bosons on a classical computer is prohibitively time-consuming for anything but the lowest truncations. The 
LSH operators and their properties may not be any more intuitive than Clebsch-Gordon coefficients, but having fully general, closed-form expressions for them allows them to be used right away on a classical computer -- and at a much faster speed than what is afforded by its parent Schwinger-boson formulation. As the full-fledged quantum simulation of SU(3) LGTs beyond 1+1 dimensions remains to be achieved, reliance on classical computing and producing benchmarks for the theory is essential and is indeed the need of the hour.
This work marks an important step towards these goals by providing explicit representations for LSH operators on higher-dimensional lattices and a ready-to-use numerical companion code for practical usefulness.

Work is in progress to extend the results of this work beyond a single site, furnishing a complete Hamiltonian matrix for an SU(3) lattice Yang-Mills theory, including a concrete numerical example. This shall be reported as part of this series in the near future. The authors believe this series would serve as fundamental building blocks for quantum simulating the dynamics of QCD and for benchmarking the results of quantum simulation, as attempted in the first large-scale demonstration of quantum simulation of LSH dynamics in Ref.~\cite{Ilcic:2026cac}.

\section{Acknowledgments}
The authors would like to thank Ivan Burbano and Anthony Ciavarella for insightful conversations at various points throughout this work.
Work by J.R.S.~was supported by the U.S. Department of Energy (DOE), Office of Science under contract DE-AC02-05CH11231, partially through Quantum Information Science Enabled Discovery (QuantISED) for High Energy Physics (KA2401032).
S.V.K.~acknowledges support by the U.S. DOE, Office of Science, Office of Nuclear Physics, InQubator for Quantum Simulation (IQuS) (award no.~DE-SC0020970), and by the DOE QuantISED program through the theory consortium ``Intersections of QIS and Theoretical Particle Physics'' at Fermilab (Fermilab subcontract no.~666484).
S.V.K.~further acknowledges the support from the Department of Physics and the College of Arts and Sciences at the University of Washington.
Research of I.R.~is supported by the  OPERA award (FR/SCM/11-Dec-2020/PHY) from BITS-Pilani, partially by the Start-up Research Grant (SRG/2022/000972) 
from Anusandhan National Research Foundation (ANRF), India and the cross-discipline research fund (C1/23/185) from BITS-Pilani. This work is part of progress made by QC4HEP working group.
A.N.~has been partially supported by the Start-up Research Grant (SRG/2022/000972) from ANRF, India received by I.R. A.N.~acknowledges the support from TCS Foundation via TCS RSP Cycle19. 

\bibliography{part-2.bib}

\end{document}